\documentclass[aps,prl,twocolumn]{revtex4}
\usepackage{graphicx}  
\usepackage{bm}        
\usepackage{amssymb}   
\usepackage{float}

\begin{document}
\title{Steady state coherence in a qubit is incompatible with a quantum map}
\author{Hans C. Fogedby}
\email{fogedby@phys.au.dk}
\affiliation{Department of Physics and Astronomy
\\
University of Aarhus, Ny Munkegade
\\
8000 Aarhus C, Denmark}
\begin{abstract}
We consider the recent proposal of steady state coherences in a single
qubit in the case of a composite system-bath interaction. Based on a 
field theoretical approach we reanalyse the issue within a Redfield 
description. We find that the Redfield approach in accordance with 
a recent proposal yields steady state coherences but also violates
the properties of a quantum map yielding negative populations. 
The issue is resolved by applying the Lindblad equation which is 
in accordance with a proper quantum map. The Lindblad equation, 
however, also implies the absence of steady state coherence. 
We conclude  that steady state coherence in a a qubit is 
incompatible with a quantum map.
 \end{abstract}
\maketitle

 {\bf Introduction:}
There is a strong current interest in open quantum systems in the context of
quantum thermodynamics \cite{Kosloff19,Rivas20,Soret22,Klatzow19}, 
quantum entanglement \cite{Streltsov17,Aolita15,Hu18,Paneru20,Horodecki09} 
and quantum computers \cite{Nandhini22}. An important issue is decoherence 
\cite{Roszak15,Ignatyuk22,Schlosshauer19} that is the loss of 
coherence due to the interaction of an open quantum system with the
environment or bath. In the case of for example entangled qubits essential for 
quantum computing maintaining coherence is thus an essential requirement.

Initiating an open quantum system in a coherent state, i.e., with off-diagonal
elements in the associated reduced density matrix, the coupling to the 
environment will result in a transient decoherence yielding a steady state
density operator with only diagonal elements corresponding to populations.
It is therefore an interesting issue whether it is possible to design an intrinsic 
system-bath interaction yielding coherence in the steady long time state, i.e.,
the generation of off-diagonal terms in the steady state density matrix.

In recent work by Guarnieri et al. \cite{Guarnieri18,Guarnieri21}, see also
\cite{Roman21,Guarnieri20}, it has been proposed that a composed 
system-bath interaction can give rise to steady state coherences in a qubit 
system. In this  work two cases are considered 1) A single  qubit subject to 
the composite  system-bath interaction of the dipole form 
$f_1\sigma^z(B^\dagger+B) + f_2\sigma^x(B+B^\dagger)$ and 2) invoking
the rotating wave approximation (RWA) the form $f_1\sigma^z(B^\dagger+B) 
+ f_2(\sigma^+B+\sigma^-B^\dagger)$. Here the Pauli matrices 
$(\sigma^x,\sigma^y,\sigma^z)$,
$\sigma^x=\sigma^++\sigma^-$, $\sigma^y=-i(\sigma^+-\sigma^-)$
 characterise the qubit states and $(B^\dagger, B)$ are the bath operators.
 
 The approach in \cite{Guarnieri18} is based on the Redfield master equation
 which is known not to represent a proper quantum map \cite{Breuer06}. 
 Whereas the Redfield equation preserves the trace of the reduced density 
 operator and thus the  conservation of probability it does not guarantee 
 positive probabilities and can in certain cases give rise to unphysical  negative  
 populations, see e.g. \cite{Fogedby24}.  Since the evolution of  a qubit subject 
 to a composite interaction should correspond to a quantum map 
 in order to represent the correct physics, it is clear that a Redfield 
 analysis could be inadequate and should be replaced by a treatment based on 
 the Lindblad equation \cite{Chrus17,Manzano20} which is known to represent 
 a proper quantum map. In the present letter we address this issue.
  
 The letter is composed of five parts. In i) we summarise a recent
 field theoretical approach to open quantum systems, in ii) we consider 
 the Redfield approach, in iii) we present the Lindblad approach, iv)  is devoted 
 to a discussion and v) to a conclusion.
  
 {\bf Field Theory:}
 The field theoretical approach to open quantum systems developed in
 \cite{Fogedby22}, see also \cite{Fogedby26}, is based on the quantum map $T$ for the
 evolution of the reduced density operator $\rho$ for an open quantum
 system, that is $\rho(t)=T\rho(0)$ \cite{Gorini76,Lax00,Lidar01}. For a 
 multi-oscillator bath  $T$ satisfies the Dyson equation  $T=T^0+T^0MT$, 
 incorporating secular effects. Here $T^0$ describes the  free propagation and 
 the kernel $M$  can be determined by an expansion  in powers of the 
 system-bath interaction $H_{SB}$ in terms of  Feynman-type diagrams. 
 In turn the  Dyson equation implies the general  (non markovian) master 
 equation
\begin{eqnarray}
\frac{d\rho(t)}{dt}=-i[H_S,\rho(t)]+\int dt'M(t,t')\rho(t'),
 \label{gen-master}
\end{eqnarray}
 where $H_S$ is the system Hamiltonian. The kernel $M$ incorporating 
 the coupling to the bath is in general complex, $M=K+i\Delta$, where 
 $K$ is the dissipative component and $\Delta$ an energy shift which 
 can be incorporated in $H_S$. In the energy basis of $H_S$, 
 $H_S|n\rangle=E_n|n\rangle$, $E_{pp'}=E_{p}-E_{p'}$,
 and invoking the condensed matter quasi-particle approximation, 
 corresponding to the  separation of time scales, we obtain the Markov 
 master equation
\begin{eqnarray}
\frac{d\rho(t)_{pp'}}{dt}=
-iE_{pp'}\rho(t)_{pp'}+ \sum_{qq'}M_{pp',qq'}\rho(t)_{qq'}.
\label{master-time}
\end{eqnarray}
Consequently, the steady state density operator $\rho^{st}_{pp'}$ is
determined by the set of linear equations
\begin{eqnarray}
-iE_{pp'}\rho^{st}_{pp'}+\sum_{qq'}M_{pp',qq'}\rho^{st}_{qq'}=0.
\label{master}
\end{eqnarray}
 Populations are given by the diagonal elements $\rho^{st}_{pp}$, 
 coherences by $\rho^{st}_{p\neq p'}$.  In the case where  
 $M_{pp,q\neq q'}=0$ the  populations form a block. Likewise,  for  
 $M_{pp',qq}=0$ the coherences  form an independent block. 
 Otherwise, the coherences interfere with  the populations. We note that 
 the trace condition $\sum_p\rho^{st}_{pp}=1$ corresponds  to 
 $\sum_pM_{pp,qq'}=0$ which by inspection  implies $\det{M}=0$,
 i.e., a steady state solution of  $-i[H_S,\rho^{st}] +M\rho^{st}=0$. 

{\bf Redfield approach:}
 In the energy representation the Redfield kernel in the Born approximation
 is given by \cite{Redfield65,Breuer06,Fogedby22,Purkayastha23}
\begin{eqnarray}
M^{Red}_{pp',qq'}=
&&-i\delta_{p'q'}\sum_{\alpha\beta l}\int\frac{d\omega}{2\pi}
\frac{S_{pl}^\alpha S_{lq}^\beta
D^{\alpha\beta}(\omega)}{E_{ql}-\omega+i\epsilon}
\nonumber
\\
&&-i\delta_{pq}\sum_{\alpha\beta l }
\int\frac{d\omega}{2\pi}\frac{S_{q'l}^\alpha S_{lp'}^\beta
D^{\alpha\beta}(-\omega)}{E_{lq'}-\omega+i\epsilon}
\nonumber
\\
&&+i\sum_{\alpha\beta }
\int\frac{d\omega}{2\pi}\frac{S_{pq}^\beta S_{q'p'}^\alpha 
D^{\alpha\beta}(-\omega)}{E_{p'q'}-\omega+i\epsilon}
\nonumber
\\
&&+i\sum_{\alpha\beta }
\int\frac{d\omega}{2\pi}\frac{S_{pq}^\beta S_{q'p'}^\alpha 
D^{\alpha\beta}(\omega)}{E_{qp}-\omega+i\epsilon},
\label{redfield}
\end{eqnarray}
where by inspection $\sum_pM^{Red}_{pp,qq'}=0$. Here $S^\alpha$ 
are the system operators and $D^{\alpha\beta}(\omega)$ the bath 
correlations.

Considering the dipole case and using the assignment in \cite{Guarnieri18} 
we choose the qubit Hamiltonian $H_S=(\omega_0/2)\sigma^z$, 
the bath Hamiltonian $H_B=\sum_k \omega_kb_k^\dagger b_k$, 
where $b_k$ is the Bose operator, the bath operator 
$E=\sum_k\lambda_k(b^\dagger_k+b_k)$, the composite system-bath 
coupling $H_{SB}=(f_1\sigma^z + f_2\sigma^x)E$ and bath 
correlation $D(\omega)=\text{Tr}_B(\rho_B EE)(\omega)$, where 
$\rho_B$ is the bath density operator. We introduce 
$S=f_1\sigma^z+f_2(\sigma^++\sigma^-)$, $E$, and $H_{SB}=SE$.
Also, $D(\omega)=g(\omega)(1+n(\omega))\eta(\omega)+
g(-\omega)n(-\omega)\eta(-\omega)$, where $g(\omega)$ is the 
spectral function and $n(\omega)=1/(\exp(\omega/T)-1)$ the 
Planck distribution, $T$ the bath temperature, and $\eta(\omega)$ 
the step function.

Inserting the non vanishing matrix elements $S_{++}=f_1$, $S_{--}=-f_1$ 
and $S_{+-}=S_{-+}=f_2$, the dissipative kernel has the $4\times 4$ 
matrix structure 
\begin{eqnarray}K^{Red}=
\left(\begin{array}{cc}
K_{pop} & K_{int}^{(a)} 
\\
 K_{int}^{(b)} &  K_{coh}
\end{array}\right),
\label{red-matrix}
\end{eqnarray}
where the population block is given by
\begin{eqnarray}
K_{pop}=
f_2^2g_0\left(\begin{array}{cc}
-1-n_0 & n_0 
\\
 1+n_0 &  -n_0
\end{array}\right),
\label{red-pop}
\end{eqnarray}
and the coherence block takes the form
\begin{eqnarray}
&&K_{coh}=
\nonumber
\\
&&-\frac{1}{2}f_2^2g_0\left(\begin{array}{cc}
1+2n_0 & 0
\\
 0&1+2n_0
\end{array}\right) 
-2f_1^2\left(\begin{array}{cc}
D_0 & 0
\\
 0&D_0
\end{array}\right).~~~~
\label{red-coh}
\end{eqnarray}
We have, moreover, introduced the interference blocks
\begin{eqnarray}
 &&K_{int}^{(a)}=
f_1f_2\left(\begin{array}{cc}
~~D_0& ~~D_0 
\\
 -D_0 &  -D_0
\end{array}\right),
\label{red-intera}
\\
&& K_{int}^{(b)}=
f_1f_2g_0\left(\begin{array}{cc}
1+n_0 &-n_0 
\\
1+n_0 &-n_0
\end{array}\right).
\label{red-interb}
\end{eqnarray}
We have used the notation $g_0=g(\omega_0)$, $n_0=n(\omega_0)$ 
and $D_0=D(0)=\lim_{\omega\to 0}g(\omega)T/\omega $.

The energy shift kernel has the matrix structure
\begin{eqnarray}\Delta^{Red}=
\left(\begin{array}{cc}
0 &0 
\\
 \Delta_{int} &  \Delta_{coh}
\end{array}\right),
\label{red-e-matrix}
\end{eqnarray}
with coherence and interference blocks
\begin{eqnarray}
&&\Delta_{coh}=
f_2^2\left(\begin{array}{cc}
-\Delta &-\Delta
\\
 ~~\Delta&~~\Delta
\end{array}\right),
\label{red-e-coh}
\\
&&\Delta_{int}=
f_1f_2\left(\begin{array}{cc}
~~\Delta_- &~~\Delta_+
\\
-\Delta_-& -\Delta_+
\end{array}\right),
 \label{red-e-int}
\end{eqnarray}
where $\Delta$ and $\Delta_\pm$ are the principal value integrals
\begin{eqnarray}
&&\Delta=\text{P}
\int\frac{d\omega}{2\pi}D(\omega)
\Big(\frac{1}{\omega+\omega_0}-
\frac{1}{\omega-\omega_0}\Big),
 \label{red-shift}
\\
&&\Delta_\pm=
2\text{P}\int\frac{d\omega}{2\pi}D(\omega)
\Big(\frac{1}{\omega\pm\omega_0}-
\frac{1}{\omega}\Big);
 \label{red-shift-pm}
\end{eqnarray}
note that $2\Delta = \Delta_+-\Delta_-$. Finally, introducing the energy matrix
\begin{eqnarray}
E=
\left(\begin{array}{cc}
0&0
\\
0&E_0
\end{array}\right),
E_0=
\left(\begin{array}{cc}
-i\omega_0&0
\\
0&+i\omega_0
\end{array}\right),
 \label{red-E}
\end{eqnarray}
the steady state density matrix $\rho^{st}$ is determined by the four 
component homogeneous linear system
\begin{eqnarray}
(E+K^{Red}+ i\Delta^{Red})\rho^{st}=0,
\label{red-lin}
\end{eqnarray}
implementing the trace condition $\rho^{st}_{++}+ \rho^{st}_{--}=1$ by a 
final normalisation.
Inspection shows that $\det(E+M^{Red})=0$, yielding a finite steady 
state density matrix $\rho^{st}_{pp'}$. In  \cite{Guarnieri18}  the trace condition
is incorporated and the inhomogeneus linear system is solved in the 
Bloch representation $\rho^{st}=(1/2)(I+\bar v\bar\sigma)$ 
yielding  $v^x,v^y,v^z$.
Noting that $\rho^{st}_{-+}=\rho^{st}_{+-}$ we find the expressions
\begin{eqnarray}
&&\rho^{st}_{++}=\frac{(\omega_0+2f_2^2\Delta)g_0n_0+2f_1^2D_0\Delta_+}
{(\omega_0+2f_2^2\Delta)g_0(1+2n_0)+4f_1^2D_0\Delta},
\label{red-rpp}
\\
&&\rho^{st}_{+-}=\frac{f_1f_2(g_0n_0\Delta_-+g_0(1+n_0)\Delta_+)}
{(\omega_0+2f_2^2\Delta)g_0(1+2n_0)+4f_1^2D_0\Delta},
\label{red-rpm}
\\
&&\rho^{st}_{--}=\frac{(\omega_0+2f_2^2\Delta)g_0(1+n_0)-2f_1^2D_0\Delta_-}
{(\omega_0+2f_2^2\Delta)g_0(1+2n_0)+4f_1^2D_0\Delta}.
\label{red-rmm}
\end{eqnarray}
These expressions are in agreement with the results in \cite{Guarnieri18}
where a detailed analysis is presented.

 {\bf Lindblad approach:}
  As shown in previous work \cite{Fogedby22,Fogedby26} the Lindblad equation can 
  be obtained  from the Redfield equation by simply applying a selection rule 
  on the Born level ensuring local energy conservation. This procedure avoids 
  applying the usual RWA \cite{Breuer06}.  The Lindblad kernel thus takes the 
  form \cite{Lindblad76,Manzano20}
\begin{eqnarray}
&&M^{Lin}_{pp',qq'}=
\nonumber
\\
&&-i\delta_{p'q'}\delta(E_{pq})\sum_{\alpha\beta l}
\int\frac{d\omega}{2\pi}
\frac{S_{pl}^\alpha S_{lq}^\beta
D^{\alpha\beta}(\omega)}{E_{ql}-\omega+i\epsilon}
\nonumber
\\
&&-i\delta_{pq}\delta(E_{p'q'})\sum_{\alpha\beta l }
\int\frac{d\omega}{2\pi}\frac{S_{q'l}^\alpha S_{lp'}^\beta
D^{\alpha\beta}(-\omega)}{E_{lq'}-\omega+i\epsilon}
\nonumber
\\
&&+i\delta(E_{pq}-E_{p'q'})\sum_{\alpha\beta }
\int\frac{d\omega}{2\pi}\frac{S_{pq}^\beta S_{q'p'}^\alpha 
D^{\alpha\beta}(-\omega)}{E_{p'q'}-\omega+i\epsilon}
\nonumber
\\
&&+i\delta(E_{pq}-E_{p'q'})\sum_{\alpha\beta }
\int\frac{d\omega}{2\pi}\frac{S_{pq}^\beta S_{q'p'}^\alpha 
D^{\alpha\beta}(\omega)}{E_{qp}-\omega+i\epsilon}.
\label{lindblad}
\end{eqnarray}
 In the non degenerate qubit case we have $\delta(E_{pq})=\delta_{pq}$ 
 and $\delta(E_{p'q'})=\delta_{p'q'}$. With the assignment in the
 case of a composite interaction as discussed in  \cite{Guarnieri18}
 and above in the Redfield case we obtain the dissipative kernel
\begin{eqnarray}K^{Lin}=
\left(\begin{array}{cc}
K_{pop} & 0 
\\
 0 &  K_{coh}
\end{array}\right),
\label{lin-matrix}
\end{eqnarray}
where the population and coherence blocks again are given
by (\ref{red-pop}) and (\ref{red-coh}), i.e. the Redfield blocks.
 
However, for  the energy shift we have
\begin{eqnarray}\Delta^{Lin}=
\left(\begin{array}{cc}
0 &0 
\\
0 &  \Delta_{coh}
\end{array}\right),
\label{lin-shift-matrix}
\end{eqnarray}
with coherence block
\begin{eqnarray}
\Delta_{coh}=
f_2^2\left(\begin{array}{cc}
-\Delta &0
\\
0&+ \Delta
\end{array}\right).
\label{lin-coh}
 \end{eqnarray}
We note that for $f_1=0$, i.e., the non-composite case the Redfield and
Lindblad kernels are identical. The difference appears in the case $f_1\neq 0$,

 {\bf Discussion:}
 1) In the {\em Redfield approach} to the composite interaction discussed in 
 \cite{Guarnieri18,Guarnieri21} we have the expressions 
 (\ref{red-rpp}-\ref{red-rmm}) for the populations $\rho^{st}_{++}$, 
 $\rho^{st}_{--}$  and the coherence term $\rho^{st}_{+-}$. For $f_1=0$ 
 the composite coupling is turned off and we obtain the usual result 
 $\rho^{st}_{++}=n_0/(1+2n_0)$, $\rho^{st}_{--}=(1+n_0)/(1+2n_0)$, i.e., 
 the equilibration of the qubit with the bath at temperature $T$ \cite{Breuer06}. 
 The coherences, $\rho_{+-}(t)\propto\exp(-i\omega_0 t-\Gamma_1 t)$, 
 $\Gamma_1=(1/2)f_2^2g_0(1+2n_0)$, decay not contributing to the 
 steady state \cite{Breuer06}. Also, only the resonant frequency $\omega_0$
 enters in the Planck factor $n_0=n(\omega_0)$, the expressions for the 
 populations being independent of the spectral strength $g_0=g(\omega_0)$. 
 In the composite case  for $f_1\neq 0$ both the splitting $\omega_0$ and 
 the shifts $\Delta$ and $\Delta_\pm$ enter in populations and induced 
 coherences. There is, moreover, a strong dependence on the low frequency 
 behaviour of the spectral function $g(\omega)$. This is due to the fact that 
 the dephasing interaction $f_1$ couples to the totality of bath modes.
  
 Assuming a spectral function of the  generic form 
 $g(\omega)=g\omega^{s}\exp(-\omega/\omega_c)$ characterised by the
 strength $g$, the parameter $s$ and the high frequency cut-off $\omega_c$ 
 and noting that $n(\omega)\sim T/\omega$ for small $\omega$, the parameter
 $D_0=\lim_{\omega\to 0}g(\omega)T/\omega$ is given by 
 $D_0=gT\omega^{s-1}$ for small $\omega$. In the ohmic case for
 $s=1$ we have $D_0=gT$. In the super-ohmic case for $s>1$ we obtain
 $ D_0=0$, yielding $\rho^{st}_{++}=n_0/(1+2n_0)$, 
 $\rho^{st}_{--}=(1+n_0)/(1+2n_0)$, and 
 $\rho^{st}_{+-}=f_1f_2(n_0\Delta_-+(1+n_0)\Delta_+)/
 ((\omega_0+2f_2^2\Delta)(1+2n_0))$. In the more interesting
 sub-ohmic case for $s<1$ we infer $D_0\to\infty$ and we obtain 
$\rho^{st}_{++}=\Delta_+/2\Delta$, $\rho^{st}_{--}=-\Delta_-/2\Delta$,
and $\rho^{st}_{+-}=0$, i.e., the absence of coherences.

For $f_2=0$ we obtain the dephasing model  \cite{Breuer06,Palma96}. 
$\sigma^z$  is a constant of motion and the populations are unchanged 
from their initial preparation at $t=0$, i.e., $\rho^{st}_{pp}= \rho_{pp}(0)$. 
From (\ref{red-coh}) we infer that the coherences decay according to  
$\rho_{+-}(t)\propto\exp(-i\omega_0 t-\Gamma_2 t)$, where
$\Gamma_2=2f_1^2D_0$. In the ohmic case for $s=1$ we have
$D_0=gT$, i.e., $\Gamma_2=2f_1^2gT$. This result is in accordance with 
the exact expression for the time dependent damping,
$\int d\omega\exp(-\omega/\omega_c)\coth(\omega/2T)(1-\cos\omega t)$, in 
the long time limit, corresponding to the limit of small $\omega$ in the integral, 
see e.g. \cite{Palma96}.

2) In the {\em Lindblad approach} by inspection of (\ref{lin-matrix}) we note the
absence of the interference blocks (\ref{red-intera}) and (\ref{red-interb})
depending on $f_1f_2$. This is a direct result of applying the selection rule,
corresponding to energy conservation at the Born level which implies that 
the matrix elements $M^{lin}_{+-,pp}$, $M^{lin}_{-+,pp}$, 
$M^{lin}_{pp,+-}$, $M^{lin}_{pp,-+}$, $M^{lin}_{+-,-+}$, and 
$M^{lin}_{-+,+-}$ all vanish identically. The selection rule implies the 
Lindblad master equation without applying the customary RWA  
\cite{Breuer06}. We note that $\det(K_{pop})=0$ entails steady
state populations, whereas $\det(K_{coh})\neq 0$ implies 
absence of coherence in the steady state. In effect, unlike the 
Redfield case in the composite case the Lindblad approach 
precludes the presence of coherence in the steady state.

\begin{figure}[H]
\includegraphics[scale=0.46]{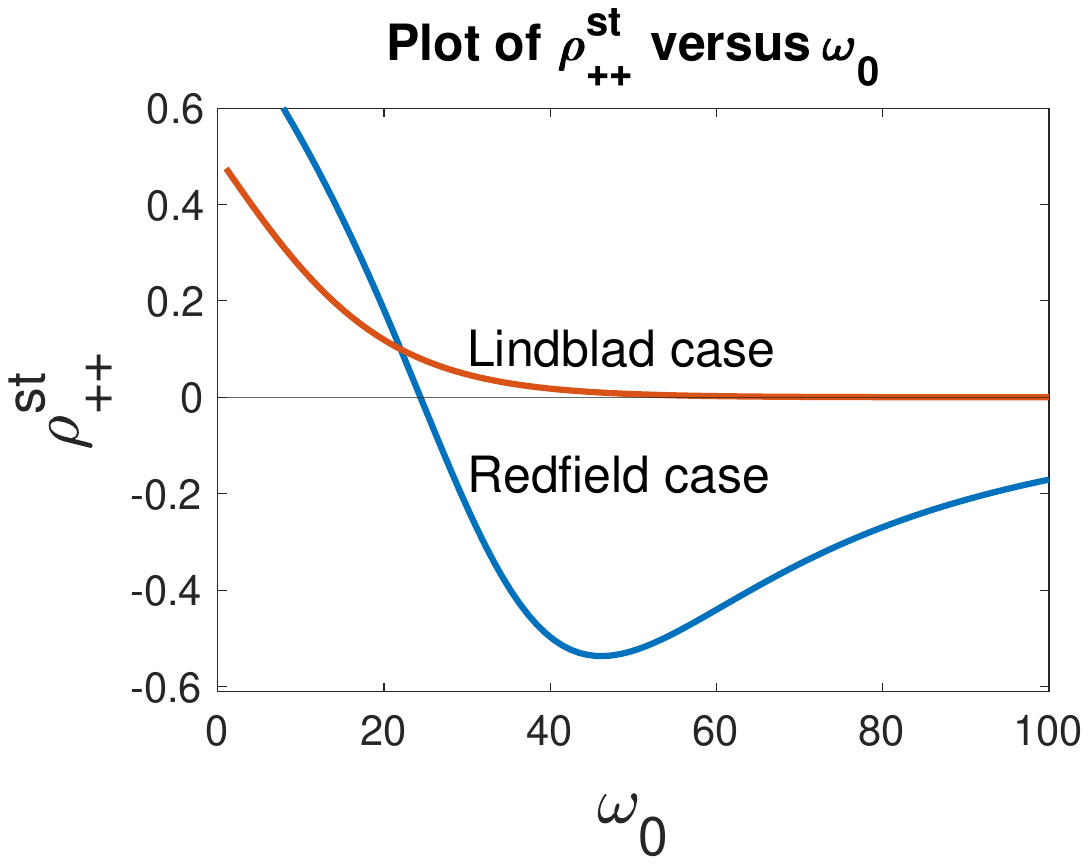}
\vspace*{-3.8 cm}
\caption{\label{fig} Plot of the population $\rho^{st}_{++}$ versus the
qubit energy splitting $\omega_0$ in the Redfield and Lindblad cases.
Temperature $T=10$, cut off frequancy $\omega_c=100$, ohmic case for
$s=1$, and spectral strength $g=1$. The Redfield case (blue) for 
$f_1=1$ and  $f_2=1$. The Lindblad case (red) for $f_1=0$ and $f_2=1$.
The Redfield case yields negative population $\rho_{++}<0$  in the range 
$\omega_0\gtrsim 25$ in violation of a quantum map.}
\end{figure}

In Fig. \ref{fig} we have plotted the population in the upper level
$\rho^{st}_{++}$ as a function of the qubit splitting $\omega_0$
in the Redfield case (blue) and the Lindblad case (red).
We have chosen the parameters $T=10$, $\omega_c=100$,
$s=1$ and $g=1$. For large $\omega_0$ the population $\rho^{st}_{++}\to 0$.
In the Lindblad case we approach $\rho^{st}_{++}=0$ through
positive values in accordance with the properties of a quantum map. 
However, in the Redfield case the population assumes unphysical 
negative values in the range $\omega_0\gtrsim 25$ in violation of the 
properties of a quantum map \cite{Gorini76,Lidar01,Breuer06}. 

 {\bf Conclusion:}
 In this letter we have discussed the issue of steady state coherence.
 In the work of Guarnieri et al. \cite{Guarnieri18} applied to a single
 qubit it is proposed that a composite interaction of the form 
 $(f_1\sigma^z+f_2\sigma^x)\sum_k\lambda_k(b_k^\dagger+b_k)$
 can generate coherence in the steady state, i.e., $\rho^{st}_{+-}\neq 0$.
 Summarising the approach in \cite{Guarnieri18}, we show that
 this is indeed the case using the Redfield master equation.
 However, several peculiarities appear in the expressions
 for the steady state density matrix; in particular a strong dependence
 on the low frequency behaviour of the spectral function. A more
 problematic issue is the appearance of negative populations.
 The last issue is in violation of the properties of a  quantum map which 
 must apply to the evolution of the density matrix. We note that the
 quantum map requirement for or open quantum  systems in the 
 Markov limit is equivalent to the requirement of unitarity in  closed 
 quantum systems.
 
 The issue is completely resolved by applying the Lindblad master
 equation which ensures a proper quantum map. The
 Lindblad approach shows that steady state coherence is
 incompatible with a quantum map. Since both the Redfield
 and Lindblad equations are derived in the second order
 Born approximation, the above "no go theorem" strictly 
 speaking only holds to second order in the system-bath coupling,
 but it seems reasonable that this might be a general result
 for a single qubit system.
 \vspace*{-3.5cm}
 
\begin{thebibliography}{30}
\expandafter\ifx\csname natexlab\endcsname\relax\def\natexlab#1{#1}\fi
\expandafter\ifx\csname bibnamefont\endcsname\relax
  \def\bibnamefont#1{#1}\fi
\expandafter\ifx\csname bibfnamefont\endcsname\relax
  \def\bibfnamefont#1{#1}\fi
\expandafter\ifx\csname citenamefont\endcsname\relax
  \def\citenamefont#1{#1}\fi
\expandafter\ifx\csname url\endcsname\relax
  \def\url#1{\texttt{#1}}\fi
\expandafter\ifx\csname urlprefix\endcsname\relax\def\urlprefix{URL }\fi
\providecommand{\bibinfo}[2]{#2}
\providecommand{\eprint}[2][]{\url{#2}}

\bibitem[{\citenamefont{Guarnieri et~al.}(2018)\citenamefont{Guarnieri, Kolar,
  and Filip}}]{Guarnieri18}
\bibinfo{author}{\bibfnamefont{G.}~\bibnamefont{Guarnieri}},
  \bibinfo{author}{\bibfnamefont{M.}~\bibnamefont{Kolar}}, \bibnamefont{and}
  \bibinfo{author}{\bibfnamefont{R.}~\bibnamefont{Filip}},
  \bibinfo{journal}{Phys. Rev. Lett.} \textbf{\bibinfo{volume}{121}},
  \bibinfo{pages}{070401} (\bibinfo{year}{2018}).

\bibitem[{\citenamefont{Kosloff and Levy}(2019)}]{Kosloff19}
\bibinfo{author}{\bibfnamefont{R.}~\bibnamefont{Kosloff}} \bibnamefont{and}
  \bibinfo{author}{\bibfnamefont{A.}~\bibnamefont{Levy}}, \bibinfo{journal}{J.
  Chem. Phys.} \textbf{\bibinfo{volume}{150}}, \bibinfo{pages}{204105}
  (\bibinfo{year}{2019}).

\bibitem[{\citenamefont{Rivas}(2020)}]{Rivas20}
\bibinfo{author}{\bibfnamefont{A.}~\bibnamefont{Rivas}},
  \bibinfo{journal}{Phys. Rev. Lett.} \textbf{\bibinfo{volume}{124}},
  \bibinfo{pages}{160601} (\bibinfo{year}{2020}).

\bibitem[{\citenamefont{Soret et~al.}(2022)\citenamefont{Soret, Cavina, and
  Esposito}}]{Soret22}
\bibinfo{author}{\bibfnamefont{A.}~\bibnamefont{Soret}},
  \bibinfo{author}{\bibfnamefont{V.}~\bibnamefont{Cavina}}, \bibnamefont{and}
  \bibinfo{author}{\bibfnamefont{M.}~\bibnamefont{Esposito}},
  \bibinfo{journal}{Phys. Rev. A} \textbf{\bibinfo{volume}{106}},
  \bibinfo{pages}{062209} (\bibinfo{year}{2022}).

\bibitem[{\citenamefont{Klatzow et~al.}(2022)\citenamefont{Klatzow, Becker,
  Ledingham, Weinzetl, Kaczmarek, Saunders, Nunn, Walmsley, Uzdin, and
  Poem}}]{Klatzow19}
\bibinfo{author}{\bibfnamefont{J.}~\bibnamefont{Klatzow}},
  \bibinfo{author}{\bibfnamefont{J.~N.} \bibnamefont{Becker}},
  \bibinfo{author}{\bibfnamefont{P.~M.} \bibnamefont{Ledingham}},
  \bibinfo{author}{\bibfnamefont{C.}~\bibnamefont{Weinzetl}},
  \bibinfo{author}{\bibfnamefont{K.~T.} \bibnamefont{Kaczmarek}},
  \bibinfo{author}{\bibfnamefont{D.~J.} \bibnamefont{Saunders}},
  \bibinfo{author}{\bibfnamefont{J.}~\bibnamefont{Nunn}},
  \bibinfo{author}{\bibfnamefont{I.~A.} \bibnamefont{Walmsley}},
  \bibinfo{author}{\bibfnamefont{R.}~\bibnamefont{Uzdin}}, \bibnamefont{and}
  \bibinfo{author}{\bibfnamefont{E.}~\bibnamefont{Poem}},
  \bibinfo{journal}{Phys. Rev. Lett.} \textbf{\bibinfo{volume}{122}},
  \bibinfo{pages}{110601} (\bibinfo{year}{2022}).

\bibitem[{\citenamefont{Streltsov et~al.}(2017)\citenamefont{Streltsov, Adesso,
  and Plenio}}]{Streltsov17}
\bibinfo{author}{\bibfnamefont{A.}~\bibnamefont{Streltsov}},
  \bibinfo{author}{\bibfnamefont{G.}~\bibnamefont{Adesso}}, \bibnamefont{and}
  \bibinfo{author}{\bibfnamefont{M.~B.} \bibnamefont{Plenio}},
  \bibinfo{journal}{Rev. Mod. Phys.} \textbf{\bibinfo{volume}{89}},
  \bibinfo{pages}{041003} (\bibinfo{year}{2017}).

\bibitem[{\citenamefont{Aolita et~al.}(2015)\citenamefont{Aolita, de~Melo, and
  Davidovich}}]{Aolita15}
\bibinfo{author}{\bibfnamefont{L.}~\bibnamefont{Aolita}},
  \bibinfo{author}{\bibfnamefont{F.}~\bibnamefont{de~Melo}}, \bibnamefont{and}
  \bibinfo{author}{\bibfnamefont{L.}~\bibnamefont{Davidovich}},
  \bibinfo{journal}{Rep. Prog. Phys.} \textbf{\bibinfo{volume}{78}},
  \bibinfo{pages}{042001} (\bibinfo{year}{2015}).

\bibitem[{\citenamefont{Hu et~al.}(2018)\citenamefont{Hu, Man, and Xia}}]{Hu18}
\bibinfo{author}{\bibfnamefont{L.-Z.} \bibnamefont{Hu}},
  \bibinfo{author}{\bibfnamefont{Z.-X.} \bibnamefont{Man}}, \bibnamefont{and}
  \bibinfo{author}{\bibfnamefont{Y.-J.} \bibnamefont{Xia}},
  \bibinfo{journal}{Quantum Inf Process} \textbf{\bibinfo{volume}{17}},
  \bibinfo{pages}{45} (\bibinfo{year}{2018}).

\bibitem[{\citenamefont{Paneru et~al.}(2020)\citenamefont{Paneru, Cohen,
  Fickler, Boyd, and Karimi}}]{Paneru20}
\bibinfo{author}{\bibfnamefont{D.}~\bibnamefont{Paneru}},
  \bibinfo{author}{\bibfnamefont{E.}~\bibnamefont{Cohen}},
  \bibinfo{author}{\bibfnamefont{R.}~\bibnamefont{Fickler}},
  \bibinfo{author}{\bibfnamefont{R.~W.} \bibnamefont{Boyd}}, \bibnamefont{and}
  \bibinfo{author}{\bibfnamefont{E.}~\bibnamefont{Karimi}},
  \bibinfo{journal}{Rep. Prog. Phys.} \textbf{\bibinfo{volume}{83}},
  \bibinfo{pages}{064001} (\bibinfo{year}{2020}).

\bibitem[{\citenamefont{snd P.~Horodecki et~al.}(2009)\citenamefont{snd
  P.~Horodecki, Horodecki, and Horodecki}}]{Horodecki09}
\bibinfo{author}{\bibfnamefont{R.~H.} \bibnamefont{snd P.~Horodecki}},
  \bibinfo{author}{\bibfnamefont{M.}~\bibnamefont{Horodecki}},
  \bibnamefont{and}
  \bibinfo{author}{\bibfnamefont{K.}~\bibnamefont{Horodecki}},
  \bibinfo{journal}{Rev. Mod. Phys.} \textbf{\bibinfo{volume}{81}},
  \bibinfo{pages}{865} (\bibinfo{year}{2009}).

\bibitem[{\citenamefont{Nandhini et~al.}(2022)\citenamefont{Nandhini, Singh,
  and Akash}}]{Nandhini22}
\bibinfo{author}{\bibfnamefont{S.}~\bibnamefont{Nandhini}},
  \bibinfo{author}{\bibfnamefont{H.}~\bibnamefont{Singh}}, \bibnamefont{and}
  \bibinfo{author}{\bibfnamefont{U.~N.} \bibnamefont{Akash}},
  \bibinfo{journal}{Advances in Engineering Software}
  \textbf{\bibinfo{volume}{174}}, \bibinfo{pages}{103337}
  (\bibinfo{year}{2022}).

\bibitem[{\citenamefont{Roszak et~al.}(2015)\citenamefont{Roszak, Filip, and
  Novotny}}]{Roszak15}
\bibinfo{author}{\bibfnamefont{K.}~\bibnamefont{Roszak}},
  \bibinfo{author}{\bibfnamefont{R.}~\bibnamefont{Filip}}, \bibnamefont{and}
  \bibinfo{author}{\bibfnamefont{T.}~\bibnamefont{Novotny}},
  \bibinfo{journal}{Scientific Reports} \textbf{\bibinfo{volume}{5}},
  \bibinfo{pages}{9796} (\bibinfo{year}{2015}).

\bibitem[{\citenamefont{Ignatyuk and Morozov}(2022)}]{Ignatyuk22}
\bibinfo{author}{\bibfnamefont{V.~V.} \bibnamefont{Ignatyuk}} \bibnamefont{and}
  \bibinfo{author}{\bibfnamefont{V.~G.} \bibnamefont{Morozov}},
  \bibinfo{journal}{Condensed Matter Physics} \textbf{\bibinfo{volume}{25}},
  \bibinfo{pages}{13302} (\bibinfo{year}{2022}).

\bibitem[{\citenamefont{Schlosshauer}(2019)}]{Schlosshauer19}
\bibinfo{author}{\bibfnamefont{M.~A.} \bibnamefont{Schlosshauer}},
  \bibinfo{journal}{Physics Reports} \textbf{\bibinfo{volume}{831}},
  \bibinfo{pages}{1} (\bibinfo{year}{2019}).

\bibitem[{\citenamefont{Guarnieri et~al.}(2021)\citenamefont{Guarnieri, Kolar,
  and Filip}}]{Guarnieri21}
\bibinfo{author}{\bibfnamefont{G.}~\bibnamefont{Guarnieri}},
  \bibinfo{author}{\bibfnamefont{M.}~\bibnamefont{Kolar}}, \bibnamefont{and}
  \bibinfo{author}{\bibfnamefont{R.}~\bibnamefont{Filip}},
  \bibinfo{journal}{Erratum Phys. Rev. Lett.} \textbf{\bibinfo{volume}{127}},
  \bibinfo{pages}{129901} (\bibinfo{year}{2021}).

\bibitem[{\citenamefont{Roman-Ancheyta
  et~al.}(2021)\citenamefont{Roman-Ancheyta, Kolar, Guarnieri, and
  Filip}}]{Roman21}
\bibinfo{author}{\bibfnamefont{R.}~\bibnamefont{Roman-Ancheyta}},
  \bibinfo{author}{\bibfnamefont{M.}~\bibnamefont{Kolar}},
  \bibinfo{author}{\bibfnamefont{G.}~\bibnamefont{Guarnieri}},
  \bibnamefont{and} \bibinfo{author}{\bibfnamefont{R.}~\bibnamefont{Filip}},
  \bibinfo{journal}{Phys. Rev. A} \textbf{\bibinfo{volume}{104}},
  \bibinfo{pages}{062209} (\bibinfo{year}{2021}).

\bibitem[{\citenamefont{Guarnieri et~al.}(2020)\citenamefont{Guarnieri,
  Morrone, Cakmak, Plastina, and Campbell}}]{Guarnieri20}
\bibinfo{author}{\bibfnamefont{G.}~\bibnamefont{Guarnieri}},
  \bibinfo{author}{\bibfnamefont{D.}~\bibnamefont{Morrone}},
  \bibinfo{author}{\bibfnamefont{B.}~\bibnamefont{Cakmak}},
  \bibinfo{author}{\bibfnamefont{F.}~\bibnamefont{Plastina}}, \bibnamefont{and}
  \bibinfo{author}{\bibfnamefont{S.}~\bibnamefont{Campbell}},
  \bibinfo{journal}{Physics Letters A} \textbf{\bibinfo{volume}{384}},
  \bibinfo{pages}{126576} (\bibinfo{year}{2020}).

\bibitem[{\citenamefont{Breuer and Petruccione}(2006)}]{Breuer06}
\bibinfo{author}{\bibfnamefont{H.~P.} \bibnamefont{Breuer}} \bibnamefont{and}
  \bibinfo{author}{\bibfnamefont{F.}~\bibnamefont{Petruccione}},
  \emph{\bibinfo{title}{The Theory of Open Quantum Systems}}
  (\bibinfo{publisher}{Oxford University Press}, \bibinfo{address}{Oxford},
  \bibinfo{year}{2006}).

\bibitem[{\citenamefont{Fogedby}(2024)}]{Fogedby24}
\bibinfo{author}{\bibfnamefont{H.~C.} \bibnamefont{Fogedby}},
  \bibinfo{journal}{J. Stat. Mech.} p. \bibinfo{pages}{P073102}
  (\bibinfo{year}{2024}).

\bibitem[{\citenamefont{Chruscinski and Pascazio}(2017)}]{Chrus17}
\bibinfo{author}{\bibfnamefont{D.}~\bibnamefont{Chruscinski}} \bibnamefont{and}
  \bibinfo{author}{\bibfnamefont{S.}~\bibnamefont{Pascazio}},
  \bibinfo{journal}{Open Systems and Information Dynamics}
  \textbf{\bibinfo{volume}{24}}, \bibinfo{pages}{1740001}
  (\bibinfo{year}{2017}).

\bibitem[{\citenamefont{Manzano}(2020)}]{Manzano20}
\bibinfo{author}{\bibfnamefont{D.}~\bibnamefont{Manzano}},
  \bibinfo{journal}{AIP Advances} \textbf{\bibinfo{volume}{10}},
  \bibinfo{pages}{025106} (\bibinfo{year}{2020}).

\bibitem[{\citenamefont{Fogedby}(2022)}]{Fogedby22}
\bibinfo{author}{\bibfnamefont{H.~C.} \bibnamefont{Fogedby}},
  \bibinfo{journal}{Phys. Rev. A} \textbf{\bibinfo{volume}{106}},
  \bibinfo{pages}{022205} (\bibinfo{year}{2022}).

\bibitem[{\citenamefont{Fogedby}(2026)}]{Fogedby26}
\bibinfo{author}{\bibfnamefont{H.~C.} \bibnamefont{Fogedby}},
  \bibinfo{journal}{arXiv:2602.13429 [quant-ph]}  (\bibinfo{year}{2026}).

\bibitem[{\citenamefont{Gorini et~al.}(1976)\citenamefont{Gorini, Kossakowski,
  and Sudarshan}}]{Gorini76}
\bibinfo{author}{\bibfnamefont{V.}~\bibnamefont{Gorini}},
  \bibinfo{author}{\bibfnamefont{A.}~\bibnamefont{Kossakowski}},
  \bibnamefont{and} \bibinfo{author}{\bibfnamefont{E.~C.~G.}
  \bibnamefont{Sudarshan}}, \bibinfo{journal}{J. Math. Phys.}
  \textbf{\bibinfo{volume}{17}}, \bibinfo{pages}{821} (\bibinfo{year}{1976}).

\bibitem[{\citenamefont{Lax}(2000)}]{Lax00}
\bibinfo{author}{\bibfnamefont{M.}~\bibnamefont{Lax}}, \bibinfo{journal}{Optics
  Communications} \textbf{\bibinfo{volume}{179}}, \bibinfo{pages}{463}
  (\bibinfo{year}{2000}).

\bibitem[{\citenamefont{Lidar et~al.}(2001)\citenamefont{Lidar, Bihary, and
  Whaley}}]{Lidar01}
\bibinfo{author}{\bibfnamefont{D.~A.} \bibnamefont{Lidar}},
  \bibinfo{author}{\bibfnamefont{Z.}~\bibnamefont{Bihary}}, \bibnamefont{and}
  \bibinfo{author}{\bibfnamefont{K.}~\bibnamefont{Whaley}},
  \bibinfo{journal}{Chemical Physics} \textbf{\bibinfo{volume}{268}},
  \bibinfo{pages}{35} (\bibinfo{year}{2001}).

\bibitem[{\citenamefont{Redfield}(1965)}]{Redfield65}
\bibinfo{author}{\bibfnamefont{A.~G.} \bibnamefont{Redfield}},
  \bibinfo{journal}{Advances in Magnetic and Optical Resonance}
  \textbf{\bibinfo{volume}{1}}, \bibinfo{pages}{1} (\bibinfo{year}{1965}).

\bibitem[{\citenamefont{Purkayastha and M{\o}lmer}(2023)}]{Purkayastha23}
\bibinfo{author}{\bibfnamefont{A.}~\bibnamefont{Purkayastha}} \bibnamefont{and}
  \bibinfo{author}{\bibfnamefont{K.}~\bibnamefont{M{\o}lmer}},
  \bibinfo{journal}{Phys. Rev. A} \textbf{\bibinfo{volume}{108}},
  \bibinfo{pages}{053704} (\bibinfo{year}{2023}).

\bibitem[{\citenamefont{Lindblad}(1976)}]{Lindblad76}
\bibinfo{author}{\bibfnamefont{G.}~\bibnamefont{Lindblad}},
  \bibinfo{journal}{Commun. Math. Phys.} \textbf{\bibinfo{volume}{48}},
  \bibinfo{pages}{119} (\bibinfo{year}{1976}).

\bibitem[{\citenamefont{Palma et~al.}(1996)\citenamefont{Palma, Suominen, and
  Ekert}}]{Palma96}
\bibinfo{author}{\bibfnamefont{G.~M.} \bibnamefont{Palma}},
  \bibinfo{author}{\bibfnamefont{K.}~\bibnamefont{Suominen}}, \bibnamefont{and}
  \bibinfo{author}{\bibfnamefont{A.}~\bibnamefont{Ekert}},
  \bibinfo{journal}{Proc. R. Soc. Lond. A} \textbf{\bibinfo{volume}{452}},
  \bibinfo{pages}{567} (\bibinfo{year}{1996}).

\end{thebibliography}

\end{document}